\documentstyle[12pt,epsfig]{article}

\newcommand{\vecvar}[1]{\mbox{\boldmath$#1$}}

\begin{document}

\vspace{0mm}
\begin{flushright}
December, 2000 \\
OU-HEP-373 \ \ \ \  
\end{flushright}
\vspace{-8mm}
\begin{center}
\large{\bf Solving the nucleon spin puzzle \\
based on the chiral quark soliton model}\footnote{Talk given
at 14th International Spin Physics Symposium (Spin2000), Osaka,
Japan, 16-21 Oct. 2000.}  

\vskip 5mm
M.~Wakamatsu

\vskip 5mm

{\small

{\it Department of Physics, Faculty of Science,}\\
{\it Osaka University, Toyonaka, Osaka 560, Japan}

}
\end{center}

\vskip 2mm

\begin{center}
\begin{minipage}{150mm}
%\centerline{\bf Abstract}
\begin{small}
\ \ \ \ \ An incomparable feature of the chiral quark soliton model
as compared with many other effective models like the MIT bag model
is that it can give reasonable predictions not only for quark
distributions but also for antiquark distributions. This will be
exemplified by the argument on the positivity constraint for
$\bar{u} (x) + \bar{d} (x)$ as well as the Soffer inequality
for quark and antiquark distributions. We also explain how the
model can resolve the so-called nucleon spin puzzle without
assuming large gluon polarization at the low energy scale.
\end{small}
\end{minipage}
\end{center}

\vspace{4mm}
\noindent
\begin{large}
{\bf 1. Introduction}
\end{large}
\vspace{3mm}

Undoubtedly, the EMC measurement in 1988 and the NMC
measurement in 1991 are two of the most striking findings in the
recent experimental studies of nucleon structure
functions \cite{EMC88,NMC91}.
A prominent feature of the chiral quark soliton model (CQSM)
is that it can simultaneously explain the above two big discoveries
in no need of artificial fine-tuning \cite{WY91,W92}.
What is the chiral quark soliton model, then?
First of all, it is a relativistic field theoretical model
effectively incorporating the idea of large $N_c$ QCD \cite{DPP88}.
For large enough $N_c$, a nucleon is thought to be a composite of
$N_c$ valence quarks and infinitely many Dirac sea quarks bound by
the self-consistent pion field of hedgehog shape.
After canonically quantizing the spontaneous
rotational motion of the symmetry breaking mean field configuration,
we can perform nonperturbative evaluation of any nucleon observables 
with full inclusion of valence and Dirac sea quarks \cite{WY91}.
It is this incomparable feature of the model that enables us to make a
reasonable estimation not only of quark distributions but
also of {\it antiquark} ones, as we shall show later.
Finally, but most importantly, only 1 parameter of the model was 
already fixed by low energy phenomenology, which means that we can give 
{\it parameter-free predictions} for parton distributions function
at the low renormalization scale.

\vspace{6mm}
\noindent
\begin{large}
{\bf 2. CQSM and twist-2 PDF}
\end{large}
\vspace{3mm}

For obtaining quark distribution functions, we need to evaluate
nucleon matrix elements of quark bilinear operators with light-cone
separation. By using the path integral formulation of the CQSM,
such nonlocality effects in time as well as spatial coordinates can be
treated in a consistent manner \cite{DPPPW96,WK99}.

The following novel $N_c$ dependencies follow from the theoretical
structure of the model, i.e. the mean-field approximation and the 
subsequent perturbative treatment of collective rotational
motion \cite{DPPPW96,WW93} :
\begin{eqnarray}
  u(x) + d(x) \ &\sim& \ N_c \,\,
  [ \,O(\Omega^0) \ + \ \ \ 0 \ \ \,] \ \ \ \ \sim \ \ O(N_c^1), \\
 u(x) - d(x) \ &\sim& \ N_c \,\,
 [ \, \ \ \vecvar{0} \ \ \ + \ O(\Omega^1) \,] \ \ \ \ \sim \ \ O(N_c^0), \\
 \Delta u(x) + \Delta d(x) \ &\sim& \ N_c \,\,
 [ \, \ \ \vecvar{0} \ \ \ + \ O(\Omega^1) \,] \ \ \ \ \sim \ \ O(N_c^0), \\
 \Delta u(x) - \Delta d(x) \ &\sim& \ N_c \,\,
 [ \,O(\Omega^0) \ + \ O(\Omega^1) \,]
 \ \ \sim \ \ O(N_c^1) \ + \ O(N_c^0) .
\end{eqnarray}
\noindent
Because of the peculiar spin-isospin correlation embedded in the 
hedgehog mean field, there is no leading-order $N_c$ contribution
to the isovector unpolarized distribution  as well as to the
isoscalar longitudinally polarized one, in contrast to the
other combinations. This especially means that the isoscalar or
flavor-singlet axial charge is parametrically
smaller than the isovector one, in conformity with the EMC observation.

\vspace{4mm}
\noindent
\begin{large}
{\bf 3. Numerical results}
\end{large}
\vspace{3mm}

In Fig.1, we summarize our parameter-free predictions for the twist-2
PDF at the model energy scale. We emphasize that {\it seeds} of all the
success of the model are already contained in these four figures. 
Here, the functions in the negative $x$
region should be interpreted as antiquark distributions according to
the rule \cite{DPPPW96} :
\begin{eqnarray}
 u(-x) \ \pm \ d(-x) \ \ &=& \ - \,\,
 [\,\bar{u}(x) \ \pm \bar{d}(x) \,]
 \ \ \ \ \ \ \ \ \ (0 < x < 1) , \\
 \Delta u(-x) \ \pm \ \Delta d(-x) \ &=& \ \ \  
 \Delta \bar{u}(x) \ \pm \ 
 \Delta \bar{d}(x) \ \ \ \ \ \ \ (0 < x < 1) .
\end{eqnarray}
The long-dashed curves peaked
around $x \simeq 1/3$ are the contributions of $N_c$ valence quarks,
while the dash-dotted curves represent
those of Dirac-sea quarks. The sum of these two contributions
are denoted by solid curves.

\begin{figure}[htbp] % fig 1
\centerline{\epsfig{file=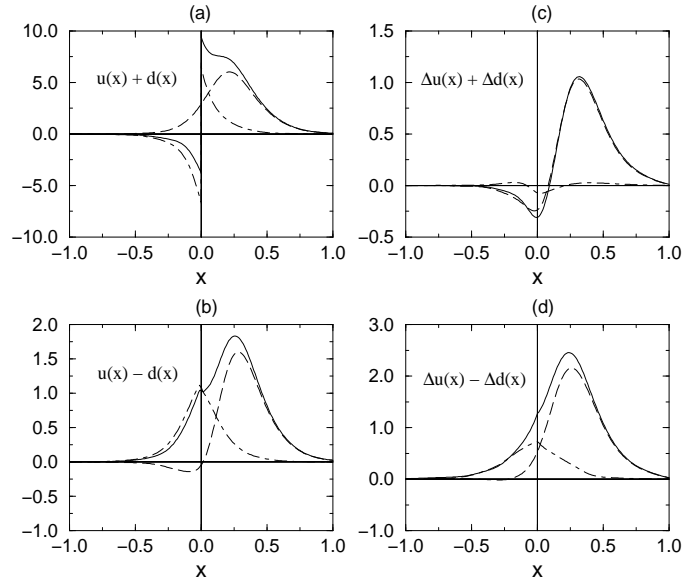,width=9.0cm,height=8.0cm}}
\begin{small}
\caption{The theoretical predictions of the CQSM for the unpolarized
distributions $u(x) + d(x)$ and $u(x) - d(x)$ as well as for the
longitudinally polarized distributions $\Delta u(x) + \Delta d(x)$
and $\Delta u(x) - \Delta d(x)$.}
\end{small}
\end{figure}

The crucial importance of the Dirac-sea contribution is most clearly
seen in the isoscalar unpolarized distribution.
Here, the ``valence-quark-only'' approximation leads to positive
$u(x) + d(x)$ in the negative $x$ region, thereby violating the
positivity of the antiquark distribution \cite{DPPPW96}.
On the other hand, if we
include the vacuum polarization of Dirac-sea quarks, the positivity
constraint for the antiquark distributions holds properly.
The effect of Dirac-sea quarks is very important also for the isovector
unpolarized distribution function \cite{WK98,PPGWW99}.
Especially interesting here is the 
fact that $u (x) - d (x) > 0$ in the negative $x$ region, which means
that $\bar{u} (x) - \bar{d} (x) < 0$ for the physical value of $x$,
just as required by the NMC measurement.
In fact, after taking account of the scale-dependence by means of
the DGLAP equation, the theory turns out to successfully explain
the NMC data for the unpolarized nucleon structure functions
$F_2^p (x)$ and $F_2^n (x)$ \cite{WW00A}.

Turning to the longitudinally polarized distributions, one observes
very different $x$ dependencies between the isoscalar and isovector ones.
One interesting feature of the isoscalar distribution is its sign change
in the small $x$ region. It has been shown that this sign change is 
just what is required by the recent experimental data for the
longitudinally polarized structure functions of the deuteron \cite{WW00A}.
Turning to the isovector distribution, we notice that the effect of 
Dirac-sea quarks has a peak of positive sign around $x \simeq 0$.
What is remarkable here is the positivity in the negative $x$ region. 
It means that anti-quark distributions are isospin asymmetric also for
the longitudinally polarized distributions \cite{WW00A}.
It is interesting to point out that some support is already given to
this unique prediction of the CQSM by several semi-phenomenological
and/or semi-theoretical analyses \cite{MY99}.

\begin{figure}[thbp] % fig 2
\centerline{\epsfig{file=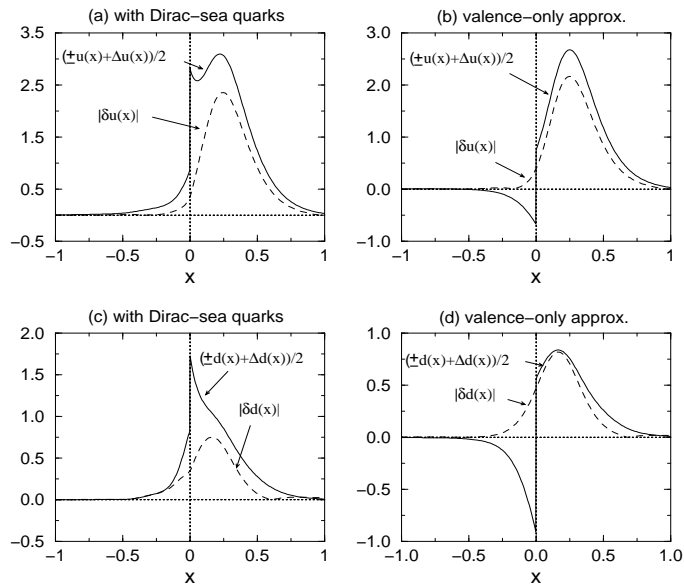,width=9.0cm,height=8.0cm}}
\begin{small}
\caption{The theoretical check of Soffer inequality.
The distributions in the negative $x$ region denote the
antiquark distributions.} 
\end{small}
\end{figure}

To complete the list of twist-2 PDF, we need another
distribution function $\delta q(x)$, usually called the transversity
distribution. It is known that this distribution function must
satisfy the so-called Soffer inequality \cite{SOF95} :
\begin{equation}
 \vert \pm \,\delta q(x) \vert \ \leq \ \frac{1}{2} \,\,
 \left( \,\pm \,q(x) \ + \ \Delta q(x) \right) \hspace{15mm}
 \left( x > 0, \ x < 0 \right) .
\end{equation}
Now the question is whether the predictions of the CQSM fulfill this
inequality or not. Fig.2 show that, if one includes the vacuum
polarization effects properly, the Soffer inequality is well satisfied
for both of $u$-quark and $d$-quark.
On the other hand, if one ignores the Dirac-sea contributions,
the Soffer inequality is badly broken for the antiquark distributions.
An important lesson learned from this observation is that the field
theoretical nature of the model, that is, the proper inclusion of
the vacuum polarization effects, plays essential roles in giving
reasonable predictions for antiquark distributions.
Another lesson is that the frequently-used saturation Ansatz of
the Soffer inequality for estimating $\delta q(x)$ is not justified.

Also noteworthy is another consequence of the soliton picture of the
nucleon. Shown in Fig.3 are the spin and the orbital angular momentum
distribution functions at the model energy scale \cite{WW00B}.
One notices that the Dirac-sea contribution to the orbital angular
momentum distribution function is sizably large and peaked around
$x \simeq 0$. Among others, large support in the negative $x$ region
suggests that seizable amount of orbital angular momentum is carried by
antiquarks. After integration over $x$, one also finds that only
about $35 \%$ of the total nucleon spin comes from the
quark spin, while the remaining $65 \%$ is due to the orbital angular
momentum of quark and antiquarks \cite{WW00B,WY91}. It is interesting
to see that the dominance of the orbital angular momentum part over
the intrinsic spin one is also indicated by the recent lattice QCD
simulation \cite{MDLMM99}.

\begin{figure}[htbp] % fig 3
\centerline{\epsfig{file=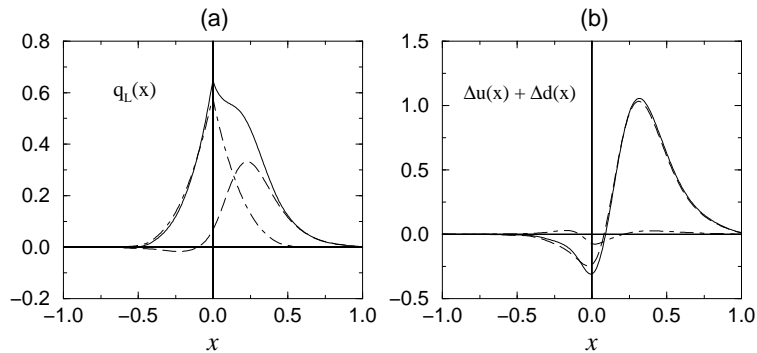,width=10.0cm,height=5.0cm}}
\begin{small}
\caption{(a) The theoretical predictions of the CQSM for the quark
and antiquark orbital angular momentum distribution functions
$q_L(x)$ and (b) the isosinglet quark polarization
$\Delta u(x) + \Delta d(x)$.
The curves have the same meaning as in Fig.1.}
\end{small}
\end{figure}

The spin and orbital angular momentum contents of the nucleon
are of course scale-dependent quantities.
We recall that, at the NLO with the gauge-invariant factorization
scheme, $\Delta \Sigma$ has a weak scale dependence mainly at low $Q^2$.
The theoretical value $\Delta \Sigma = 0.31$ obtained at
$Q^2 = 10 \,\mbox{GeV}^2$ is qualitatively consistent with the recent
SMC result, $\Delta \Sigma^{exp}_{SMC} = 0.22 \pm 0.17$ \cite{SMC98}.

\vspace{4mm}
\noindent
\begin{large}
{\bf 4. Conclusion}
\end{large}
\vspace{3mm}

In summary, an incomparable feature of the CQSM as
compared with many other effective models like the MIT bag model
is that it can give reasonable predictions also for the
{\it antiquark distribution functions} as exemplified by the argument
on the possitivity constraint for $\bar{u}(x) + \bar{d}(x)$ and
also on the Soffer inequality for antiquarks.
It has been emphasized that parton distribution functions evaluated
at the model energy scale contain all the {\it seeds} of the success of
the model in explaining existing experimental data given at the high
energy scale. It naturally explains the excess of $\bar{d}$ sea over
the $\bar{u}$ sea in the proton.
The most puzzling observation, i.e. unexpectedly small quark spin
fraction of the nucleon can also be explained in no need of
large gluon polarization at the low energy scale.

As a further unique prediction of the model, we pointed out the
possibility of large {\it isospin asymmetry} of the {\it spin-dependent
sea-quark distributions}, which seems to be a natural consequence
of the large $N_c$-counting rule, but appears inconsistent with the naive
``meson cloud convolution model''.
Then, if this large asymmetry of the
longitudinally polarized sea is experimentally established, it
would offer a strong evidence in favor of nontrivial spin-isospin
correlation imbedded in the ``large $N_c$ chiral soliton picture''
of the nucleon.

\vspace{3mm}
The talk is based on the collaborations with T.~Watabe and T.~Kubota.
\vspace{-8mm}

%
%  Reference
%
\vspace{4mm}
\setlength{\baselineskip}{5mm}

\end{document}